# Omnidirectional wave energy gimbal-based electromagnetic generator

Zhichao Jiang[1]*, Shunchao Jiang[1]
[1]School of Physical Science & Technology, Guangxi University, Nanning 530004, China
* Corresponding author email: shzhch1992@163.com

Wave energy, as a renewable energy source, is widely distributed and possesses substantial reserves. However, many existing wave energy harvesters exhibit motion constraints under irregular wave conditions, which limits their energy conversion efficiency. In this study, an omnidirectional wave energy gimbal-based electromagnetic generator (OWG-EMG) is proposed to achieve stable power generation under irregular wave motions. By integrating a three-axis gimbal mechanism with an inertial mass regulator, the proposed harvester effectively converts complex and varying wave-induced motions into relative rotational motion between gimbal frames. A planetary gear transmission is further employed to increase the rotational speed of the electromagnetic generator, enabling efficient omnidirectional wave energy harvesting. Experimental results demonstrate that the device delivers an average output power of up to 0.24 W at a frequency of 1 Hz, corresponding to an average power density of 40 W/m³. Importantly, in addition to its high energy conversion efficiency, the proposed device offers notable advantages including low cost, structural simplicity, and high reliability, highlighting its significant potential for applications in ocean energy harvesting and utilization.

## I. INTRODUCTION

Amid the escalating global energy crisis and intensifying climate change, wave energy has emerged as a clean energy source of considerable strategic importance. With its high energy density and extensive distribution, it presents a promising solution for sustainable development [1]. The development of wave energy conversion technologies will facilitate the advancement of sustainable energy systems and provide direct and stable power supplies for marine facilities [2]. However, the randomness and irregularity of wave motion result in unstable outputs from wave energy harvesters, severely constraining their power generation performance. Therefore, overcoming the multidirectional variability of waves to achieve omnidirectional, efficient, and low-cost wave energy harvesting has become a central technical challenge in this field.

In recent years, extensive efforts have been devoted to improving wave energy harvesting efficiency, primarily focusing on two aspects: power generation mechanisms and structural design of devices. In terms of generation mechanisms, research has mainly concentrated on electromagnetic generators(EMGs) and triboelectric nanogenerators (TENGs) [3–6]. TENGs have attracted considerable attention due to their flexible structural designs that can adapt to complex marine environments [7–10]. However, the inherently high output voltage and low output current of triboelectric generation lead to relatively low overall energy conversion efficiency. In contrast, electromagnetic generation offers advantages such as high output current and high energy density. Nevertheless, because it relies on rotational motion to cut magnetic flux lines, conventional TENGs face difficulties in adapting structurally to low-frequency and multidirectional wave environments [11,12]. From the perspective of structural design, most existing wave energy harvesting devices feature internal mechanisms with limited motion directions, making it challenging to stably and efficiently convert wave energy from varying directions [13–15]. To address this limitation, some studies have attempted to introduce multi-axis gimbal mechanisms. Owing to their ability to significantly enhance degrees of freedom in motion, such mechanisms are commonly integrated with gyroscopes to form three-axis gyroscopic systems. Several wave energy generation devices have adopted similar three-axis gyroscopic structures in their designs [16,17]. However, although inspired by three-axis gyroscopes, these existing approaches have not fully exploited the multidirectional motion capability of gimbal mechanisms, and thus the challenge of truly omnidirectional wave energy harvesting remains unresolved.

Herein, we propose an efficient omnidirectional wave energy gimbal-based electromagnetic generator (OWG-EMG). The device converts complex, irregular, and multidirectional wave motions into controlled relative rotations between gimbal-mounted frames. Through the integration of a planetary gear transmission mechanism, it achieves a substantial increase in the rotational speed of the EMG. In this way, efficient capture and conversion of complex wave energy are achieved. Experimental results demonstrate that the harvester delivers an average output power of 0.24 W at a frequency of 1 Hz, corresponding to an average power density of 40 W/m³. The device consists of four individual EMG units; a single EMG can charge a 0.47 F supercapacitor to 3 V within 50 s, enabling it to power various electronic devices. Importantly, in addition to its high-power generation performance, the proposed device features low cost, structural simplicity, and high reliability, demonstrating substantial potential for applications in ocean energy harvesting and utilization.

## II. STRUCTURAL DESIGN OF OWG-EMG

As illustrated in **Fig. 1**, the overall structural schematic of the OWG-EMG is presented. The device features a cylindrical configuration, with stabilizing fins mounted at the bottom to maintain stability on the sea surface and to minimize undesired self-rotation. The interior of the cylindrical shell consists of a three-axis gimbal mechanism, an inertial mass regulator, and an electromagnetic generator. Under the action of the inertial mass, wave energy is converted into relative rotational motion between the gimbal frames, thereby driving the EMG to rotate and generate electricity.

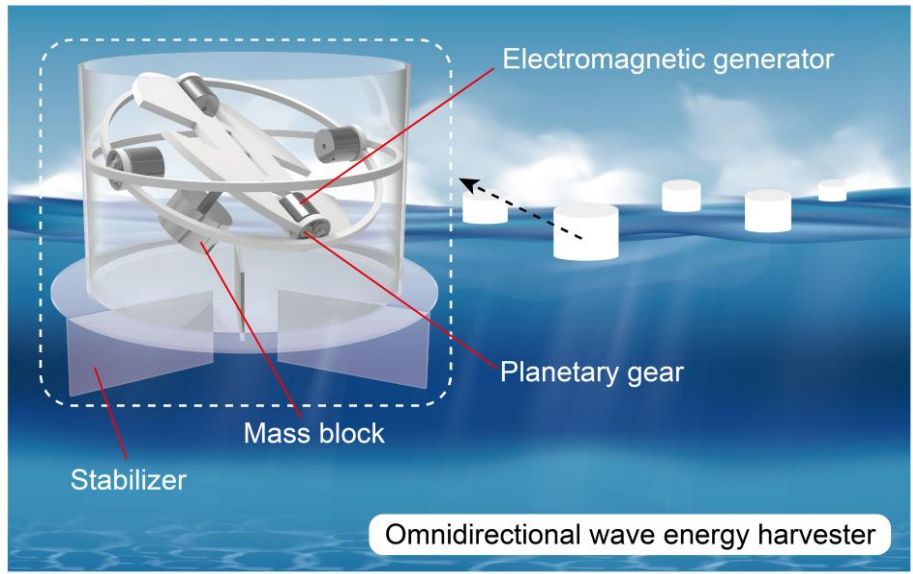

**Fig.1.** Overall structural schematic diagram of OWG-EMG.

**Fig. 2** presents a photograph of the fabricated OWG-EMG device. The outermost gimbal frame is fixed inside the cylindrical shell to support the entire gimbal mechanism. Relative rotational motion is permitted between adjacent gimbal frames. The cylindrical housing is fabricated from transparent acrylic, while the gimbal frames are manufactured via three-dimensional (3D) printing. The EMG employed in the system is a commercially available unit. Owing to the low-cost materials and straightforward fabrication process, the overall device is highly suitable for large-scale commercial production.

its unconstrained directional motion characteristics, the mechanism is capable of harvesting wave energy from multiple directions under irregular wave excitations, thereby enabling omnidirectional energy capture.

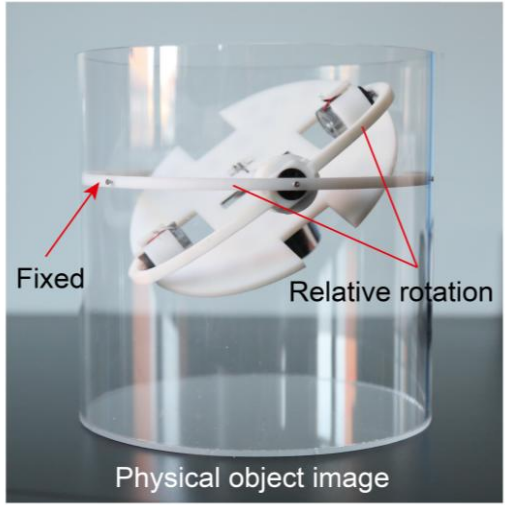

**Fig.2.** The physical object image of OWG-EMG.

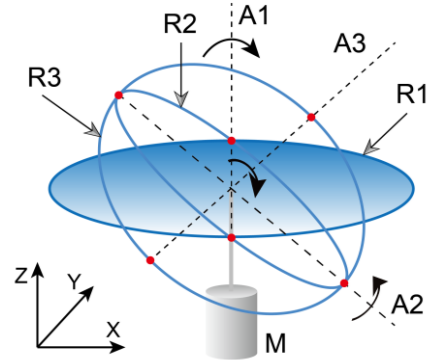

**Fig.3.** Mechanism diagram of mechanical motion.

**Fig. 3** illustrates the kinematic mechanism of the OWG-EMG. Theoretically, the system constitutes a three-axis gimbal mechanism. As shown in the figure, gimbal frame R2 serves as the intermediate layer and rotates about the central axis A1 of gimbal frame R1. Gimbal frame R3 forms the inner layer and rotates about the central axis A2 of R2. The inertial mass block M rotates about the central axis A3 of R3. These three relative rotational motions collectively establish a three-axis gimbal mechanism. Owing to

**Fig. 4** illustrates the motion states of the gimbal frames under excitations from different directions. When the central axis of gimbal frame R2 is parallel to the excitation direction (**Fig. 4a**), no relative rotation occurs between R3 and R2; under this condition, only Generators 1 and 2 are operational. When the central axis of R3 is parallel to the excitation direction (**Fig. 4b**), no relative rotation occurs between R2 and R1, and thus only Generators 3 and 4 are activated. When the central axis of either R2 or R3 is not parallel to the excitation direction (**Fig. 4c**), inertial forces induce relative rotational motion among all three gimbal frames. In this case, all four generators operate simultaneously, resulting in the maximum electrical energy output.

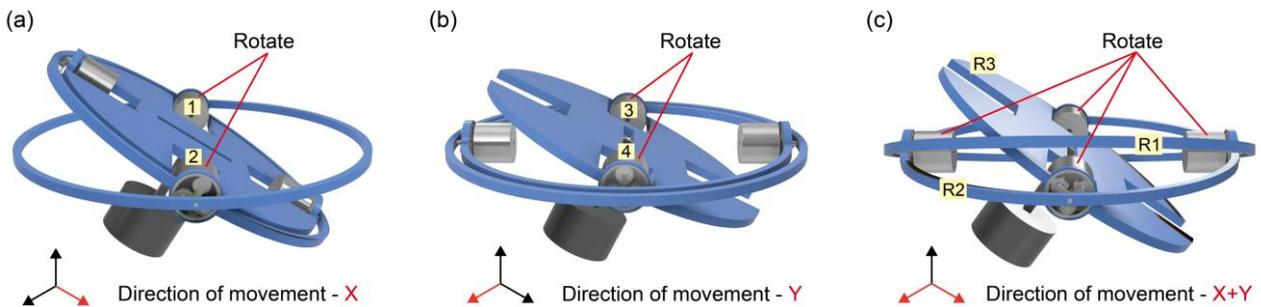

**Fig.4.** The response characteristics of the gimbal to different directions of motion. (a) Relative motion state of the gimbal in the X direction. (b) Relative motion state of the gimbal in the Y direction. (c) Relative motion state of the gimbal in the composite direction.

**Fig. 5** presents the detailed configuration of the power generation module, in which a planetary gear transmission mechanism is employed to enhance the rotational speed of the EMG. As illustrated, the EMG is rigidly coupled to the planetary gear assembly. The ring gear of the planetary gear mechanism is integrated with gimbal frame R2, while the planet carrier is fixedly connected to gimbal frame R1. When relative rotational motion occurs between R1 and R2, synchronous motion is generated between the ring gear and the planet carrier. Through the transmission effect of the planetary gear set, the low-speed relative rotation between the gimbal frames is converted into high-speed rotation of the sun gear. Consequently, the EMG achieves enhanced rotational speed and improved electrical output performance.

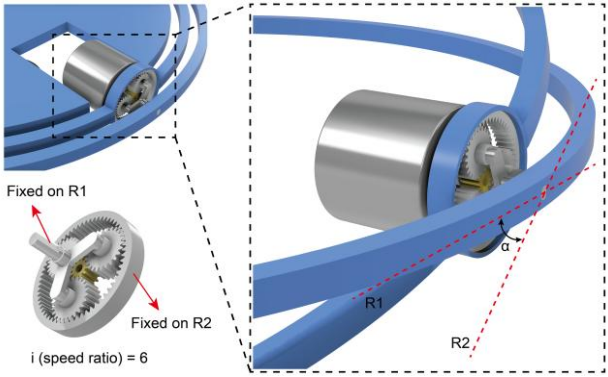

**Fig.5.** Partial schematic diagram of the power generation mechanism.

### III. Output Performance of OWG-EMG

The output performance of a single EMG was evaluated under laboratory conditions. The entire OWG-EMG device was fixed onto the sliding platform of a reciprocating linear motor to simulate wave-induced oscillations. As shown in **Fig. 6**, at an oscillation frequency of 1 Hz, a peak output voltage of 10.5 V was obtained from a single EMG. Furthermore, impedance matching measurements (**Fig. 7**) revealed that the instantaneous output power and average output power of a single EMG reached 145 mW and 60 mW, respectively, at a matched load resistance of 200 Ω. The system design ensures that at least two EMGs are operational, guaranteeing a minimum total average power output exceeding 120 mW, reaching up to 240 mW.

**Fig. 8** presents the charging current of a single EMG when charging a 0.6 F capacitor. The output terminal of the EMG is directly connected to the capacitor through a rectifier bridge, enabling direct DC charging. The measured charging current reaches up to 15 mA.

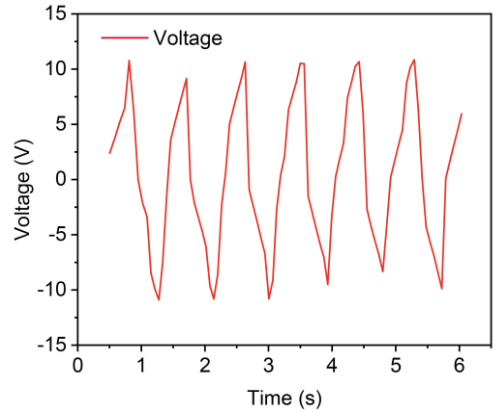

**Fig. 6.** Open-circuit voltage of a single EMG.

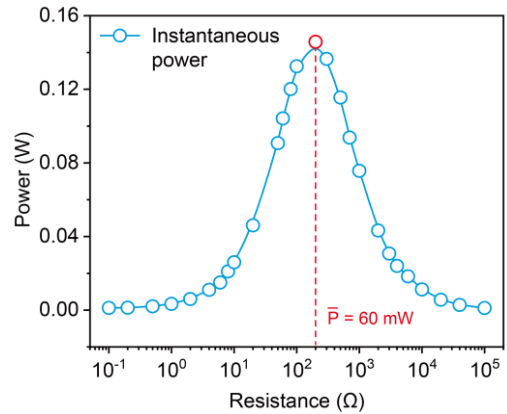

**Fig.7.** Output power of a single EMG.

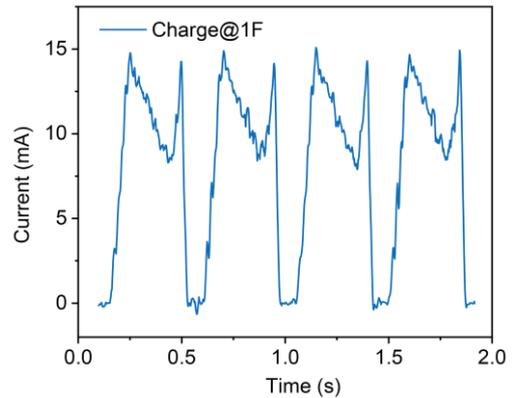

**Fig.8.** Charging current of a single EMG during the supercapacitor charging.

**Fig. 9** shows the charging curve of a single EMG charging a 0.47 F supercapacitor. As observed, the supercapacitor can be charged to 3 V within 50 s. **Fig. 10** illustrates the charging performance of a single EMG charging a 0.6 F supercapacitor. The results indicate that the voltage reaches 3 V within 150 s.

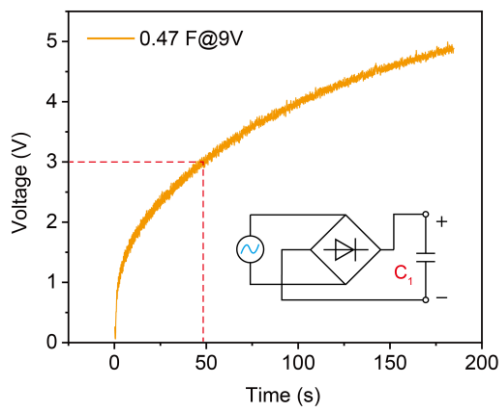

**Fig.9.** Charging curve for a 0.47 F supercapacitor.

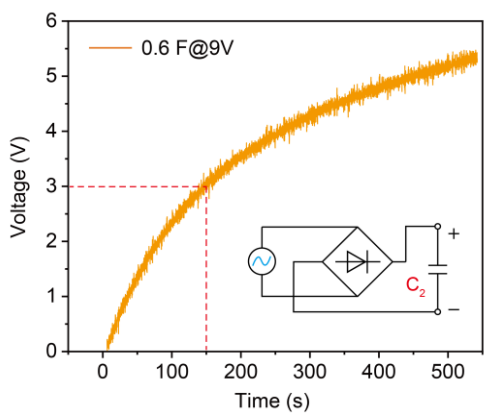

**Fig.10.** Charging curve for a 0.6 F supercapacitor.

## IV. CONCLUSION

In summary, the OWG-EMG not only demonstrates the capability for omnidirectional wave energy harvesting but also exhibits excellent power generation performance. At an excitation frequency of 1 Hz, the device achieves an average output power of up to 0.24 W, corresponding to an average power density of 40 W/m³. Notably, a single EMG can charge a 0.47 F supercapacitor to 3 V within 50 s, providing sufficient power for various electronic devices. In addition to its high energy conversion efficiency, the OWG-EMG features significant advantages, including low cost, structural simplicity, and high reliability. These merits highlight its considerable potential for practical applications in ocean energy harvesting and utilization.